\documentclass[aps,prl,reprint,superscriptaddress]{revtex4-1}
\usepackage{mathtools,xspace}
\usepackage{amssymb,amsfonts,amsthm}
\usepackage{subcaption}
\usepackage{graphicx}
\usepackage{colonequals}
\usepackage[usenames,dvipsnames]{xcolor}
\usepackage{url}
\usepackage[outline]{contour}
\contourlength{1.2pt}
\usepackage{framed}
\usepackage{calc}
\usepackage[clockwise]{rotating}
\usepackage{fancyhdr}
\usepackage{multirow}
\usepackage{floatpag}
\usepackage{wrapfig}
\usepackage{floatrow}
\usepackage{microtype}
\usepackage{hyperref}
\hypersetup{
 pdftitle={Active spanning trees and Schramm-Loewner evolution},
 pdfauthor={Adrien Kassel and David B. Wilson},
colorlinks=true,
linkcolor=NavyBlue,
urlcolor=Blue,
citecolor=PineGreen,
}
\usepackage{bookmark}

\newcommand{\old}[1]{}

\newtheorem*{conjecture}{Conjecture}

\newcommand{\G}{\mathcal{G}}

\newcommand{\SLE}{\operatorname{SLE}}
\newcommand{\CLE}{\operatorname{CLE}}
\newcommand{\ia}{\operatorname{ia}}
\newcommand{\ea}{\operatorname{ea}}
\newcommand{\act}{\operatorname{a}}

\begin{document}
\title{Active~spanning~trees~and~Schramm--Loewner~evolution}
\author{Adrien Kassel}
\affiliation{Departement Mathematik, ETH, 8092 Z\"{u}rich, Switzerland}
\author{David B.\! Wilson}
\affiliation{Microsoft Research, Redmond, Washington 98052, USA}

\begin{abstract}
We consider the Peano curve separating a spanning tree from its dual
spanning tree on an embedded planar graph, where the tree and dual tree are
weighted by $y$ to the number of active edges, and ``active'' is in
the sense of the Tutte polynomial.  When the graph is a portion of
the square grid approximating a simply connected domain, it is known
($y=1$ and $y=1+\sqrt{2}$) or believed ($1<y<3$) that the Peano curve converges to a
space-filling $\SLE_{\kappa}$ loop, where $y=1-2\cos(4\pi/\kappa)$,
corresponding to $4<\kappa\leq 8$.
We argue that the same should hold for $0\le y<1$, which corresponds to
$8<\kappa\leq 12$.
\end{abstract}
\maketitle

\bookmark[dest=introduction]{Introduction}
\hypertarget{introduction}{}
One of the aims of statistical physics is to describe the phenomenology of phase transitions. These can be broadly classified into discontinuous or continuous transitions. For the latter, the main challenge is to derive critical exponents which govern the behavior of physical quantities at the transition point.
To that end, a popular approach is to study discrete mathematical models which are both tractable and yet complicated enough that they reflect universal features of critical phenomena. Sorting these models into universality classes independent of the details of each model is of central importance.

A family of such discrete models is the Fortuin--Kasteleyn (FK) random-cluster model~\cite{FK}, which is a probability measure on subsets of edges of a graph (i.e., a correlated percolation model), where a configuration is weighted by $p/(1-p)$ to the number of edges times $q$ to the number of connected components.
Although many quantitative features of the FK model have been predicted using
the renormalization group method,
conformal field theory, and Coulomb gas methods,
rigorous mathematical derivations are only sparse and recent.  In two dimensions, the phase transition in $p\in[0,1]$ was predicted by Baxter \cite{baxter} to be continuous for $q\le 4$ and discontinuous for $q>4$; for the square lattice, this has been
proven rigorously for $1\leq q\leq 4$ \cite{DCST} and $q\geq 26$ \cite{laanait-message-ruiz}.
On the square lattice, the critical value is expected to be the self-dual value $\sqrt{q}/(1+\sqrt{q})$, which was proven rigorously~\cite{Beffara-Duminil} for $q\ge 1$.

In 2D, a topologically equivalent way of encoding the information of such percolation models is to look at the boundaries of connected components (called clusters) which form a dense collection of loops.
There is a way to cut these loops open and connect them together to obtain a unique space-filling loop (a ``Peano curve'') which can be seen as a Markovian exploration of the FK configuration of edges. We shall describe this in greater detail below.

Conformal Field Theory (CFT) has predicted the structure of universality classes in 2D by means of representations of the Virasoro algebra (parametrized by the central charge). In a major breakthrough~\cite{Schramm:SLE}, Schramm gave a new geometrical way of apprehending these universality classes by introducing a family of random fractal curves defined in planar domains and stochastically invariant under conformal transformations: the Schramm--Loewner evolutions (SLE). The family is parametrized by a nonnegative real number~$\kappa$ accounting for the fractal dimension $\min(1+\kappa/8,2)$ of the curves \cite{rohde-schramm,Beffara}.

The SLE curves describe a one-parameter family of universality classes for planar critical models.
The loops surrounding the critical FK clusters are believed to have the same scaling limit as the $O(n)$ loop model in its dense phase where $n=\sqrt{q}$.
As explained in the expository paper~\cite{kager-nienhuis:SLE}, the general prediction for the relations between the parameter $q$ of the critical FK model, the $O(n)$ model, and the parameter $\kappa$ of the corresponding $\SLE_\kappa$ is
\begin{equation}\label{n-kappa}
\sqrt{q}=n=-2\cos(4\pi/\kappa) \,.
\end{equation}
This means that interfaces between clusters are expected to converge to $\SLE_\kappa$ where $\kappa$ is the largest solution of~\eqref{n-kappa}~\cite{Schramm-ICM}. Equivalently, the corresponding Peano curve should converge to a space-filling variant of $\SLE_\kappa$ \cite{miller-sheffield:ig4}.
This has been proven in certain cases, notably for the critical FK-Ising (i.e.\ $q=2$) interfaces by Smirnov and collaborators~\cite{CDCHKS:ising,kemppainen-smirnov}. In the limit $q\to 0$, the critical FK model becomes the uniform spanning tree, for which the Peano curve was proven by Lawler, Schramm, and Werner~\cite{LSW:UST} to be described by~$\SLE_8$.

For each $n\in[-2,2]$, the two largest solutions $\kappa_1\ge \kappa_2$ to~\eqref{n-kappa} are related by $1/\kappa_1+1/\kappa_2=1/2$. For $n\ge 0$, the $O(n)$ loop model on the honeycomb lattice (with edge-weight $\mathsf{x}$) should have (at least) two conformally invariant phases: one at the critical point $\mathsf{x}_c=1/\sqrt{2+\sqrt{2-n}}$ (dilute phase, described by $\SLE_{\kappa_2}$), and one in the supercritical regime $\mathsf{x}>\mathsf{x}_c$ (dense case, described by $\SLE_{\kappa_1}$);
see e.g.\ \cite[\S~5.6]{kager-nienhuis:SLE} and also \cite{blote-nienhuis}. In addition to the dilute and dense phases, there is also a compact (fully packed) phase corresponding to $\mathsf{x}=\infty$.  On the honeycomb lattice, it was predicted to be conformally invariant~\cite{KdGN}.  However, the scaling limit is lattice dependent and for example its critical exponents differ on the honeycomb and square lattices~\cite{BBNY} (see also \cite{jacobsen:compact}).

To date, discrete models corresponding to the universality classes of $\SLE_\kappa$ have essentially only been defined in the range
$2\leq\kappa\leq 8$.  (There was a proposal that paths within ``watersheds''
converge to $\SLE_{1.734\pm0.005}$ \cite{watersheds-sle},
but that is wrong \cite{WW,wilson:RGB}.)
It is a natural objective to look for discrete models corresponding to
other values of $\kappa$.

We propose a probabilistic model --- generalizing the uniform spanning tree model ---
which we conjecture converges to 
a form of $\SLE_\kappa$ in the range $\kappa\in(8,12]$ (and also in the dual range $\kappa\in [4/3,2)$).
 Prior to SLE, the $O(n)$ model was studied, from the point of view of CFT, in the range $-2\leq n\leq 2$ by Nienhuis~\cite{Nienhuis-82} and Cardy~\cite{Cardy}, but for $n<0$ it no longer makes sense as a probability measure on loop configurations. (The range $n<0$ corresponds to central charge $c<-2$.)  Our model gives a probabilistic representation of the critical FK model corresponding to the range $\sqrt{q}=n\geq -1$, using a combinatorial description of the Tutte polynomial of a graph, which we now review.

\medskip
\bookmark[dest=tutte]{Tutte polynomial}
\hypertarget{tutte}{\textit{\textbf{Tutte polynomial.}}}
For an arbitrary graph~$\G=(V,E)$, the Tutte polynomial is
\begin{equation}\label{subgraph-expansion}
T_\G(x,y)\colonequals
\sum_{E'\subseteq E}(x-1)^{k(E')-k(E)}(y-1)^{k(E')+|E'|-|V|}\,,
\end{equation}
where the sum is over subsets of edges $E'$,
and $k(E')$ is the number of connected components of the spanning subgraph of $\G$ with edge set $E'$.
The partition function of the FK model is a specialization of the Tutte polynomial (up to a multiplicative constant) with $x=1+q(1-p)/p$ and $y=1+p/(1-p)$
\cite{FK}.
In the following, we assume without loss of generality that the graph $\G$ is connected. Tutte's original definition is
\begin{equation}\label{tree-expansion}
T_\G(x,y) = \sum_{\text{spanning trees $t$}} x^{\ia(t)}\, y^{\ea(t)}\,,
\end{equation}
where the sum is over spanning trees $t$, $\ia(t)$ is the number of ``internally active'' edges of $t$, and $\ea(t)$ is the number of ``externally active'' edges of $t$.  The notion of active edge requires some explanation.

An edge is said to be internal with respect to a spanning tree if it is part of it, and external otherwise. The cycle formed by adding an external edge to a spanning tree is called its fundamental cycle with respect to the tree. The cut formed by deleting an internal edge is called its fundamental cut with respect to the tree.
In Tutte's definition~\cite{Tutte}, the edges come with an arbitrary order,
and whether or not an internal (resp.\ external) edge is active with respect to a spanning tree is determined by whether or not
the edge is minimal amongst all edges in its fundamental cut (resp.\ cycle) with respect to the tree.
There are other definitions of activities~\cite{Bernardi0,Bernardi,Courtiel} for which, remarkably,~\eqref{tree-expansion} always holds.

With Tutte's
original definition of activity, the equivalence of 
\eqref{subgraph-expansion} and~\eqref{tree-expansion} can
be verified using an edge contraction / deletion recursion formula.
We use a notion of activity which is defined using the
planar embedding of the graph, as discussed below.  This ``embedding''
activity was essentially described by Bernardi, except that we replace
``minimal'' with ``maximal'' in Bernardi's definition
\cite[\S~3.1, Def.~3]{Bernardi},
so that it can be understood in terms of a local
exploration process; see also Courtiel's discussion
\cite[\S~7.2]{Courtiel}.  This local exploration process is very
similar to the discrete version of Sheffield's SLE exploration tree
construction of the conformal loop ensemble
\cite[\S~2.1]{Sheffield-exploration}
which was also used in \cite{Sheffield-HCbijection}. 

Assume now that the graph $\G$ is embedded in the plane and let $\G^*$ be its dual. The medial graph of $\G$ is a $4$-valent planar graph obtained by creating a $4$-valent vertex
at the intersection of each edge with its dual;
medial graph vertices are neighbors whenever the corresponding edges of~$\G$ both share a vertex and bound the same face.

\begin{figure}[b!]
\includegraphics[width=\textwidth]{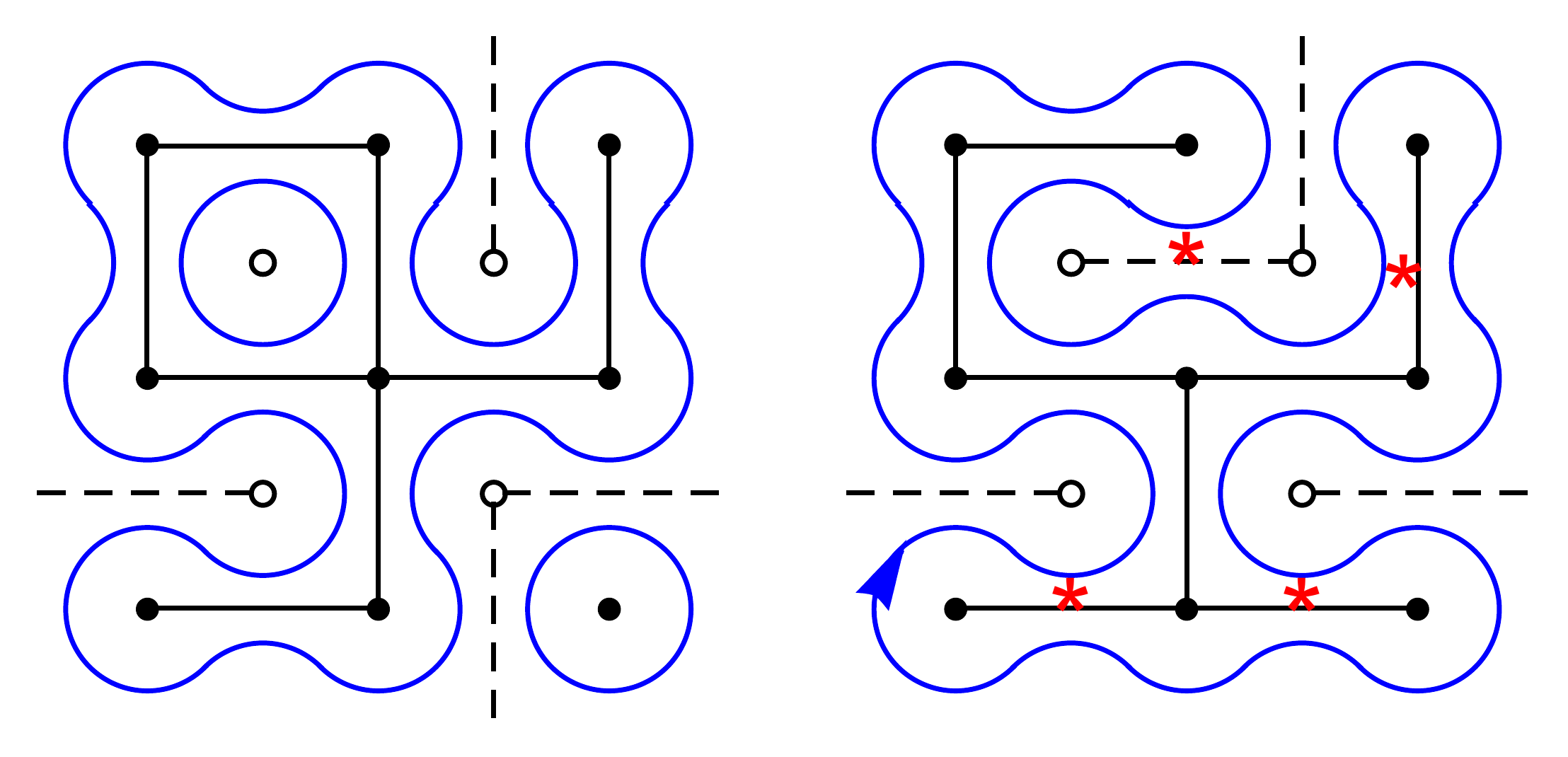}
\caption{Left: a subgraph of the $3\times 3$ square grid with free boundary conditions, together with the dual subgraph (dotted lines) of the $2\times 2$ grid with wired boundary conditions (the outer wired dual vertex is not shown), and its
 Eulerian loop representation (blue curved loops).
Right: the associated spanning tree, dual spanning tree,
and Peano curve (blue loop based at the arrowhead), along with stars marking the active edges. Each star is drawn on the side of the edge first encountered by the Peano curve, which is when the exploration processs (defined below) determines the edge to be active.
}\label{fig:tutte}
\end{figure}

The medial graph provides a way of rewriting the Tutte polynomial of a planar graph in a more symmetric way \cite{BKW,Jaeger-88}:
at any medial graph vertex, there are two ways to split the vertex into two noncrossing 2-valent vertices;
the resulting configuration is a collection of loops on the medial graph which is Eulerian (every edge of the medial graph is used exactly once).
Given a subgraph of $\G$, one can split the medial graph vertices
so that the Eulerian loops do not cross the edges of the subgraph or its dual.
This defines a mapping from subgraphs of~$\G$ to Eulerian loop configurations of its medial graph. See Fig.~\ref{fig:tutte}.

Given a subset $E'$ of primal edges, let $E'^*$ be the set of dual edges not crossing $E'$ (of cardinality $|E|-|E'|$). By Euler's formula, \eqref{subgraph-expansion} may be rewritten
\[T_{\G}(x,y)=\sum_{E'\subset E}(x-1)^{k(E')-1}(y-1)^{k(E'^*)-1}\,.\]

By the above mapping, and labelling (by ``iloop'' or ``eloop'') each Eulerian loop whether it is the exterior boundary of a primal or dual cluster, further yields
\[T_{\G}(x,y)=\sum_{\text{Eulerian loop config.}}(x-1)^{\# iloops-1}(y-1)^{\#eloops}\,,\]
where the sum is over all Eulerian loop configurations of the medial graph.

We now define an exploration procedure for connecting together the loops in an Eulerian loop configuration. Start at the middle of an edge of the medial graph and pick a direction for the exploration.  Each medial vertex is split in one of two ways when it is first encountered by the exploration process.  If one of the possible splits would disconnect the graph or dual graph, the other split is used, and that medial vertex is (embedding) active.  Otherwise, the medial vertex is split according to the Eulerian loop configuration, and the medial vertex is not active.

The exploration process produces a new Eulerian loop configuration consisting of just one loop.  This loop is the ``Peano curve'' separating a spanning tree $t$ from its dual spanning tree $t^*$. See Fig.~\ref{fig:tutte}.

For each spanning tree~$t$, there are $2^{\act(t)}$ loop configurations that map to it, where $\act(t)=\ia(t)+\ea(t)$ is the total number of active edges for $t$.  Starting from a spanning tree $t$, if we switch any $k$ internally active edges and $\ell$ externally active edges, we obtain an Eulerian loop configuration with $k+1$ internal loops and $\ell$ external loops.
Hence the weighted sum over Eulerian loop configurations can be rewritten as a sum over spanning trees each having weight $x^{\ia(t)} y^{\ea(t)}$.
This implies the equivalence of~\eqref{subgraph-expansion} and~\eqref{tree-expansion} for planar graphs.
(See also \cite{Courtiel} or \cite{Bernardi}.)

\begin{figure}[h!]
\includegraphics[width=\textwidth]{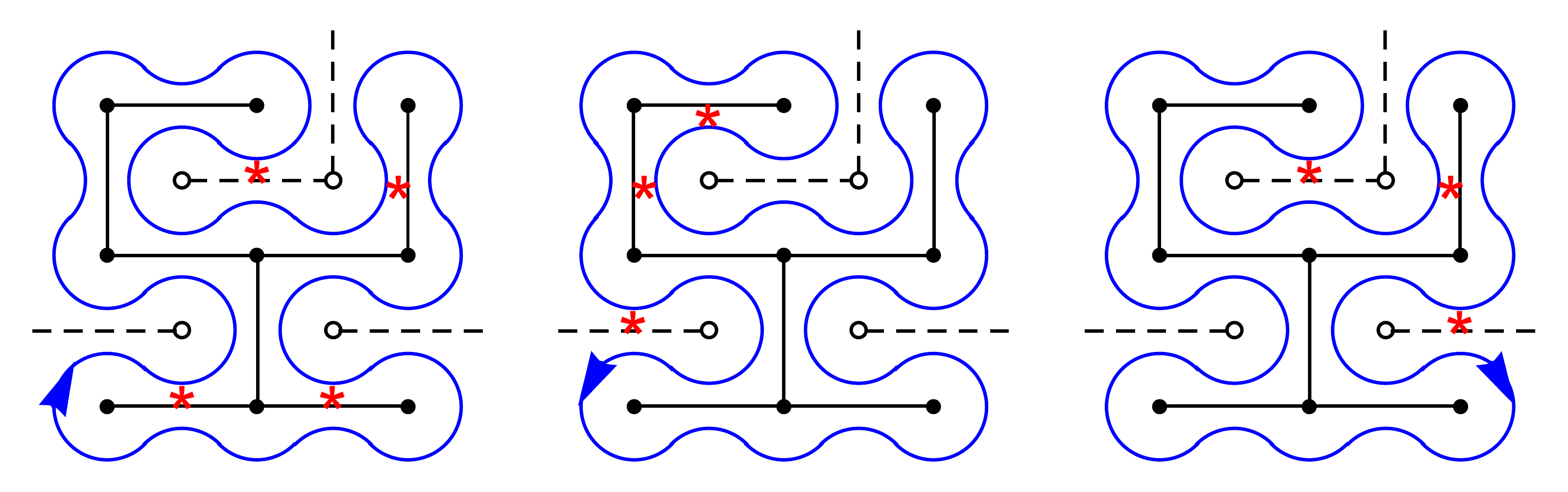}
\caption{The spanning tree~$t$ from Fig.~\ref{fig:tutte}.  When the
  Peano curve is rooted at a different location, or its orientation is
  reversed, the number of active edges $\act(t)$ can change.}
\label{fig:rooted}
\end{figure}

Note that the activity $\act(t)$ of a spanning tree depends on the starting point and direction of the exploration process,
as shown in Fig.~\ref{fig:rooted}.

\begin{figure}[t!]
\centerline{\includegraphics[width=1.65in]{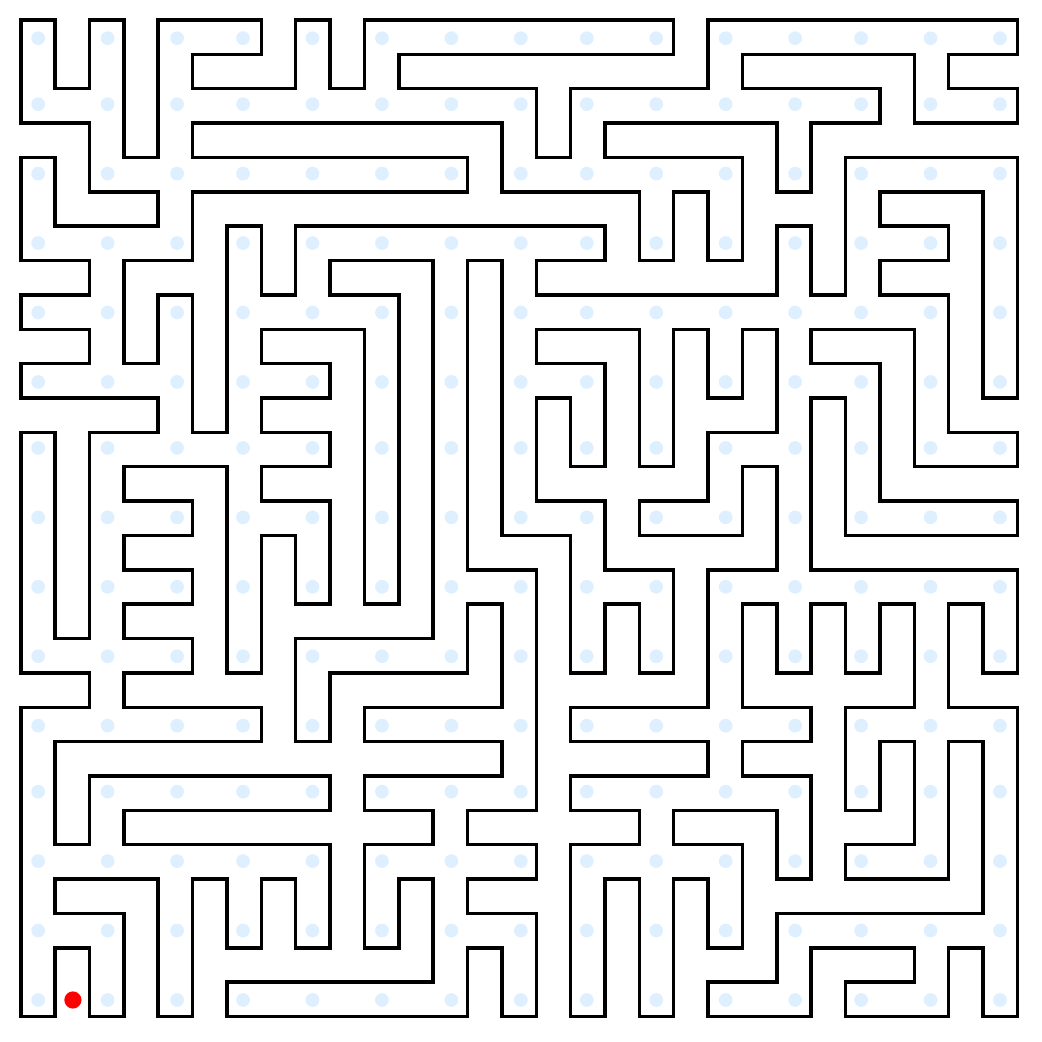}\hfill
\includegraphics[width=1.65in]{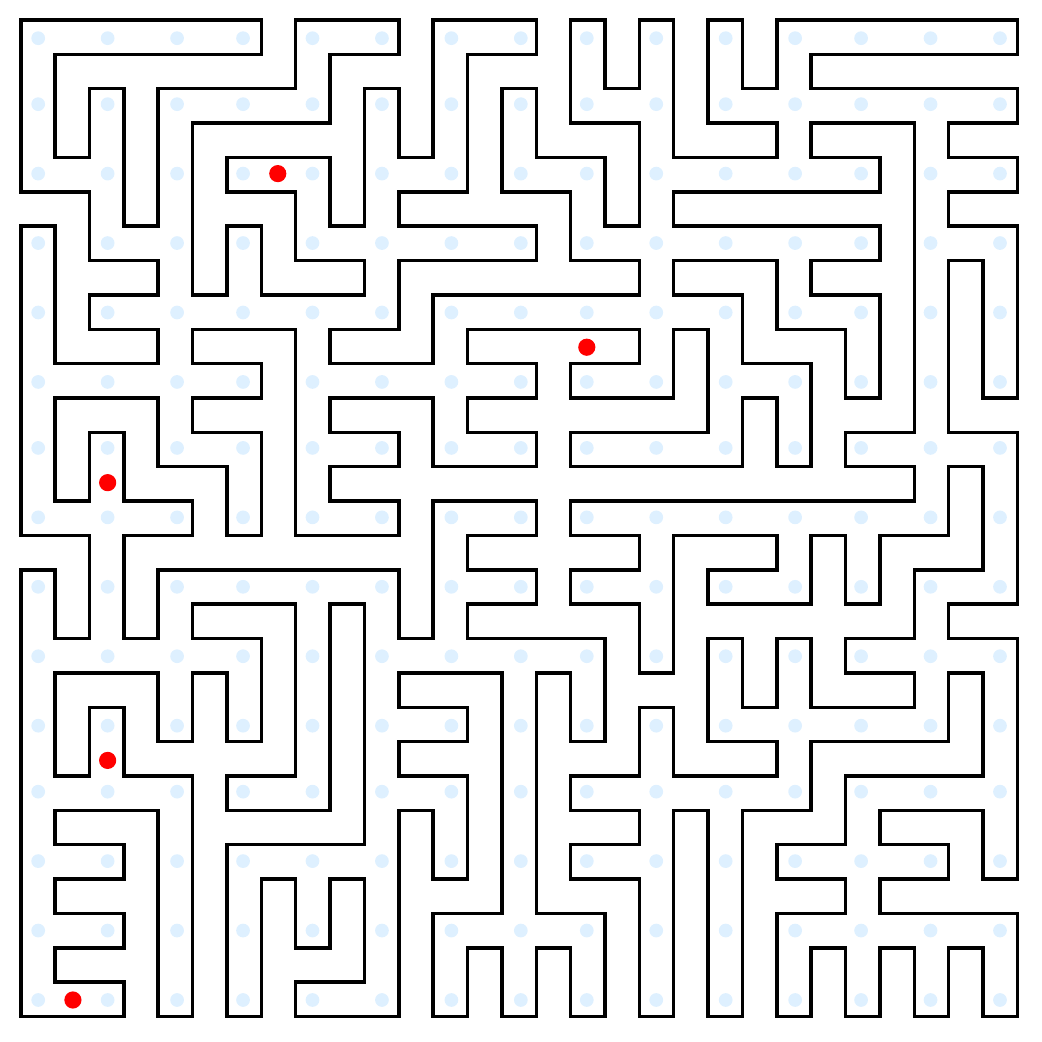}}
\vspace{8pt}
\centerline{\includegraphics[width=1.65in]{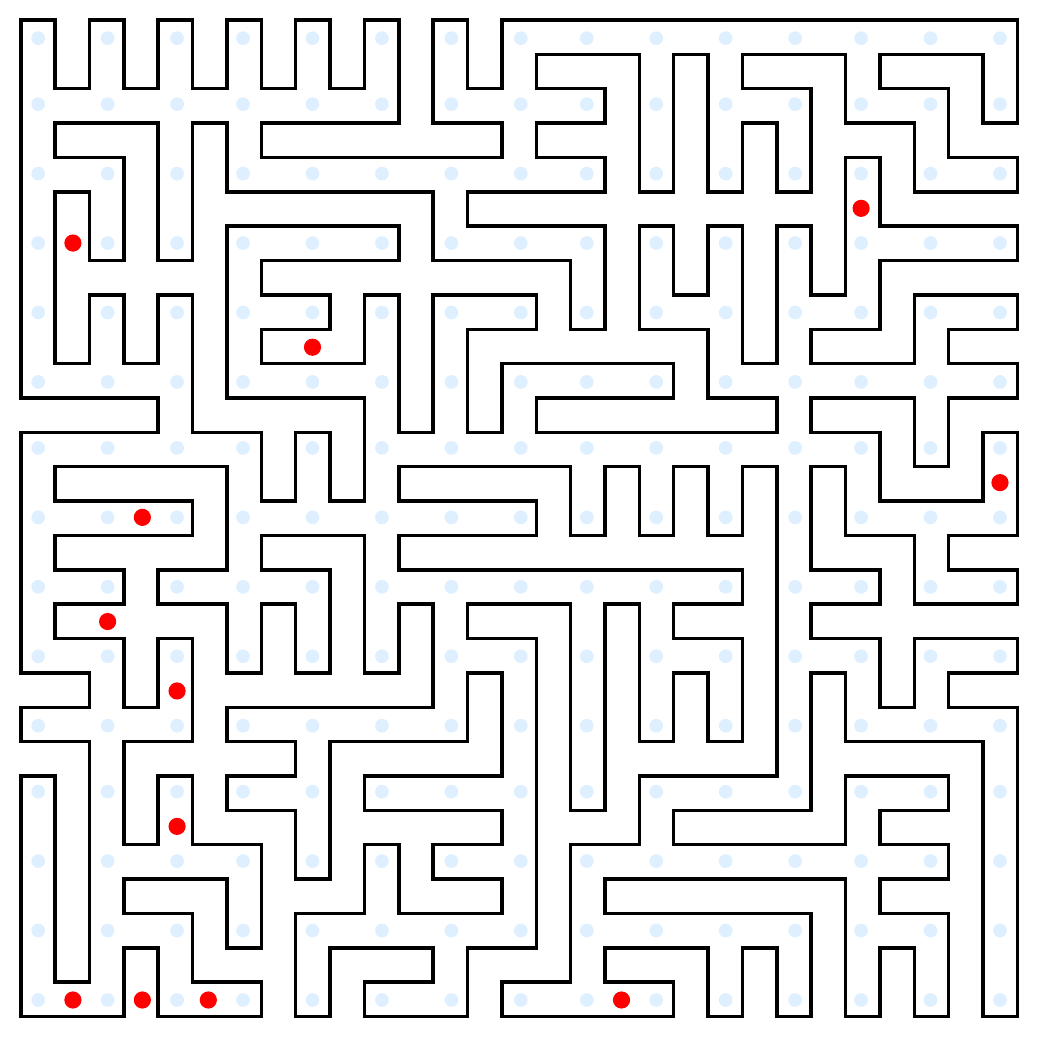} \hfill
\includegraphics[width=1.65in]{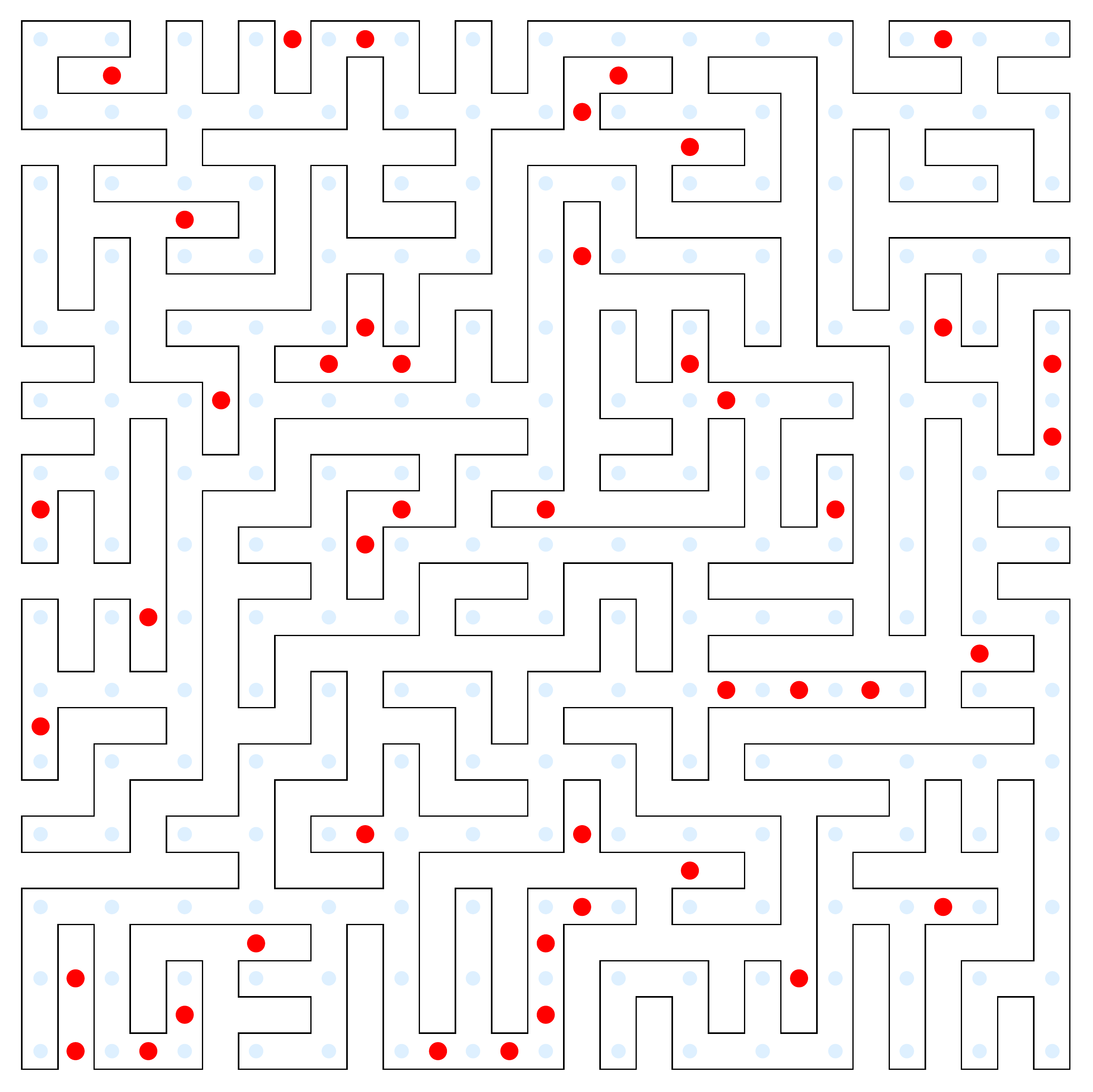}}
\caption{Active spanning trees on a $15\times 15$ grid with active edges (both internal and external) marked with red dots. Here the Peano curve starts in the lower left corner and runs clockwise (as in Fig.~\ref{fig:tutte}).
The values of $y$ are $1/64$ (upper left), $1/16$ (upper right), $1/8$ (lower left), and $1/2$ (lower right).
}\label{Samples}\label{Peano}
\end{figure}

\newcommand{\Ltlabel}{\llap{\raisebox{15pt}{$L$\hspace{4pt}}}\llap{\raisebox{5pt}{\rotatebox[origin=c]{-20}{\footnotesize\#sweeps}\hspace{57pt}}}}
\newcommand{\ytlabel}{\llap{\raisebox{45pt}{$y$\hspace{10pt}}}\llap{\raisebox{15pt}{\rotatebox[origin=c]{-20}{\footnotesize\#sweeps}\hspace{100pt}}}}
\begin{figure*}[t]
\begin{subfigure}{.195\textwidth}\includegraphics[width=\textwidth]{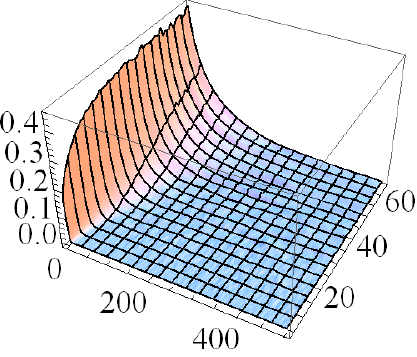}\Ltlabel\end{subfigure}
\begin{subfigure}{.195\textwidth}\includegraphics[width=\textwidth]{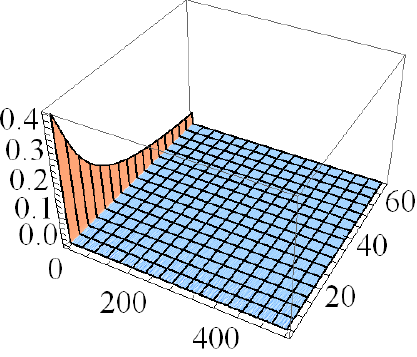}\Ltlabel\end{subfigure}
\begin{subfigure}{.195\textwidth}\includegraphics[width=\textwidth]{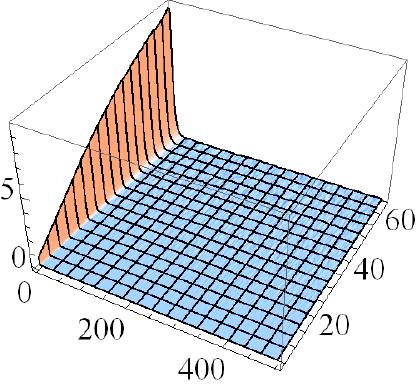}\Ltlabel\end{subfigure}
\begin{subfigure}{.195\textwidth}\includegraphics[width=\textwidth]{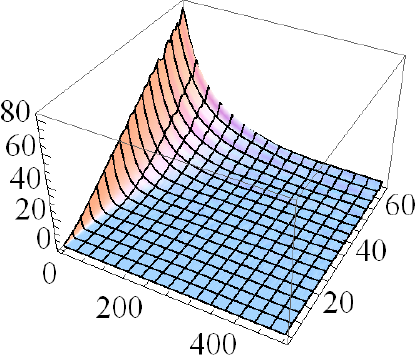}\Ltlabel\end{subfigure}
\begin{subfigure}{.195\textwidth}\includegraphics[width=\textwidth]{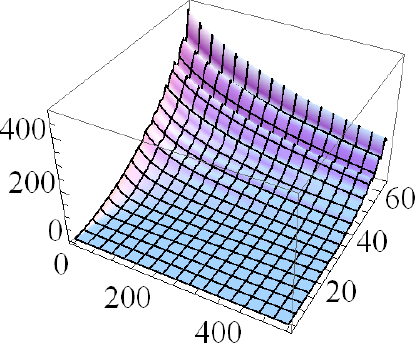}\Ltlabel\end{subfigure}
\caption{Autocovariance for the winding angle variance estimator for fixed $y$ as a function of $L$ and the number of sweeps.
Here ``sweep'' means $L^2$ independent updates of the active-tree Markov chain. From left to right,
 the values of $y$ are $1/48$, $1$, $2$, $3$, and $4$.
}
\label{fig:autocov-Lt}
\end{figure*}

\begin{figure*}[t]
\begin{tabular}{lcr}
\includegraphics[width=\textwidth/3]{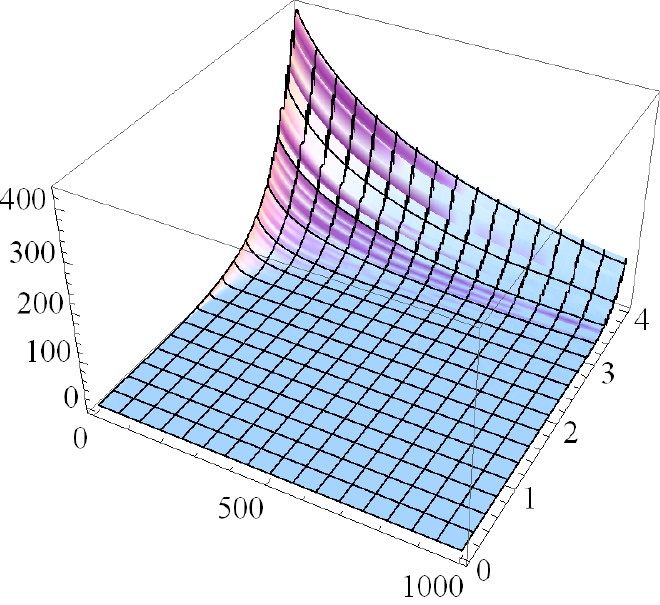}\ytlabel & 
\includegraphics[width=\textwidth/3]{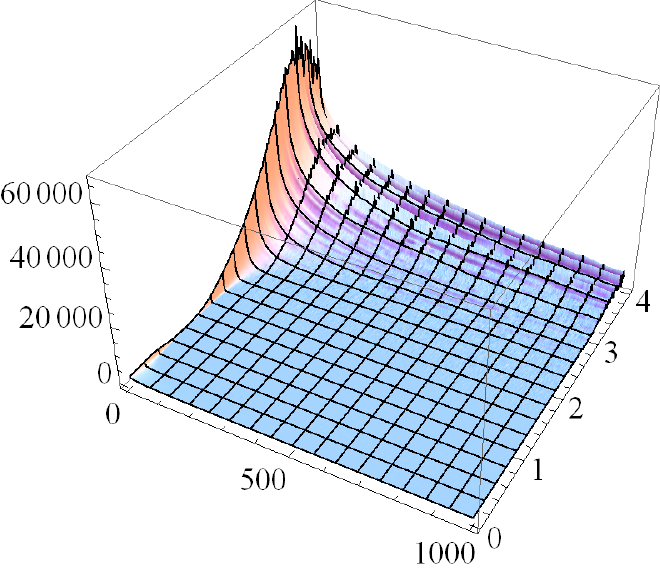}\ytlabel & 
\includegraphics[width=\textwidth/3]{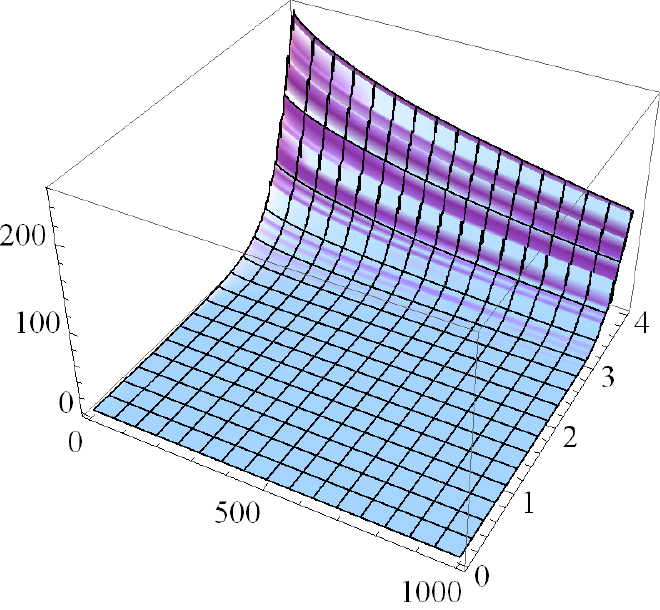}\ytlabel
\end{tabular}
\vspace{-2pt}
\caption{Autocovariance for the estimators for winding angle variance (left), outer boundary dimension (middle), and outer boundary winding angle variance (right), for $L=64$ as a function of $y$ and time.  These estimators appear to decorrelate rapidly (after a small number of sweeps of the active-tree Markov chain) for $0<y\leq 3$.}
\label{fig:autocov-yt}
\end{figure*}

\medskip
\bookmark[dest=active-tree]{Active spanning trees}
\hypertarget{active-tree}{\textit{\textbf{Active spanning trees.}}}
The exploration procedure maps any subgraph to a spanning tree.  In particular, a self-dual FK model with parameter~$q$
is mapped to a random spanning tree~$t$ with distribution proportional to $y^{\act(t)}$, where $y=n+1$,
and $\act(t)$ is the embedding activity.
We call such a tree an ``active spanning tree'' (see Fig.~\ref{Peano}).
It is analogous to Sheffield's exploration tree process
for constructing the conformal loop ensembles $\CLE_\kappa$, with
$4<\kappa\leq 8$ \cite{Sheffield-exploration}.
This analogy holds for $1\leq y$, which is the range of $y$
for which the FK interpretation is valid.

It is natural to consider random trees with similar probability distributions \eqref{tree-expansion} which do not come from the exploration of a random subgraph: active spanning trees weighted by $y^{\act(t)}$ for $0\le y<1$.
Importantly, their Peano curves satisfy the domain Markov property, even without an associated FK representation.
Note that the domain Markov property holds for the embedding activity, but not Tutte's original definition of activity.

In view of CFT predictions~\cite{Nienhuis-82,Cardy} for $n=y-1\in[-1,0)$, it is natural to assume that in the range $y<1$, the model should also exhibit conformal invariance.  Hence, if the Peano curve has a limit, satisfying both the domain Markov property and conformal invariance (and is space-filling), it must be an $\SLE_{\kappa}$ curve for some parameter $\kappa \ge 8$.  We conjecture that the relation between $n=y-1$ and $\kappa$ should still be~\eqref{n-kappa} in that range.

Before stating the conjecture, we review some background on imaginary geometry, the theory recently developped by Miller and Sheffield.
For $\kappa\leq 4$, $\SLE_\kappa$ can be constructed~\cite{dubedat:SLE-GFF,sheffield:welding,miller-sheffield:ig1} as flow lines of the vector field $e^{ih/\chi}$
where $h$ is a Gaussian free field and $\chi=2/\sqrt{\kappa}-\sqrt{\kappa}/2$.
Miller and Sheffield showed that when one glues together all possible flow lines in this field,
one obtains a continuum spanning tree whose branches are $\SLE_\kappa$ and for which the curve separating
it from its dual tree is a space-filling version of $\SLE_{\kappa'}$ for $\kappa'=16/\kappa$ \cite{miller-sheffield:ig1,miller-sheffield:ig4}. This is the so-called light cone duality.
(When $\kappa'\geq 8$, $\SLE_{\kappa'}$ is already space-filling \cite{rohde-schramm}; for $4<\kappa'<8$ Miller and Sheffield explain
how to construct space-filling $\SLE_{\kappa'}$ by splicing together ordinary $\SLE_{\kappa'}$'s \cite{miller-sheffield:ig4}.)

SLE duality for $\kappa'>8$ is slightly different than for $\kappa'\leq8$.
$\SLE_{\kappa'}$ is reversible when $\kappa'\leq 8$, but not when $\kappa'>8$, although in this case the variant $\SLE_{\kappa'}(\rho_1,\rho_2)$ with force points is reversible provided $\rho_1+\rho_2=\kappa'/2-4$ \cite{Zhan,miller-sheffield:ig4}.  The outer boundary of $\SLE_{\kappa'}$ (with or without force points) is $\SLE_{16/\kappa'}$ with suitable force points, but the formula for the forces is different when $\kappa'>8$ \cite{Zhan,miller-sheffield:ig4}.

\begin{conjecture}
The Peano curve of an active spanning tree with $0\le y<1$ converges in the scaling limit towards $\SLE_{\kappa'}$ (with force points), where
 $\kappa'=4\pi/\arccos((1-y)/2)\in(8,12]$.
\end{conjecture}

This would imply, by $\SLE$-duality~\cite{Zhan,Dubedat,miller-sheffield:ig1}, that the branches of the active tree converge to $\SLE_\kappa$ (with force points) with $\kappa=16/\kappa' \in [4/3,2)$.

For $1\le y<3$, the analogous conjecture is equivalent to the prediction for the critical FK model scaling limit~\cite{kager-nienhuis:SLE}.

In joint work with Gwynne and Miller \cite{GKMW:active-tree-map},
we prove that
in the setting of quantum gravity, where the underlying graph is itself random,
and weighted by the active tree partition function, for $y<1$
the active tree Peano curve converges to $\SLE_{\kappa'}$ in the peanosphere sense \cite{DMS:mating,miller-sheffield-peano,GHMS:covariance}.
(The case $1\leq y<3$ was proven earlier by Sheffield \cite{Sheffield-HCbijection}; see also \cite{gwynne-mao-sun,gwynne-sun-II,gwynne-sun-III}.)

The case $y=0$ corresponds to the uniform measure on spanning trees with minimal number of active edges.
Bernardi~\cite{Bernardi} constructed a bijection between these minimally active trees and bipolar orientations, where a bipolar orientation is an acyclic orientation with exactly one source and one sink
(an earlier bijection was given by Gioan and Las Vergnas \cite{gioan-vergnas} for Tutte's notion of activity).
With Bernardi's bijection, our conjecture implies that a certain bipolar Peano curve
defined in \cite{KMSW1} should converge to $\SLE_{12}$ and the branches of the tree to $\SLE_{4/3}$.
This was proven recently in the setting of quantum gravity~\cite{KMSW1}.

\begin{figure*}[t]
\begin{center}
\includegraphics[width=2.2in]{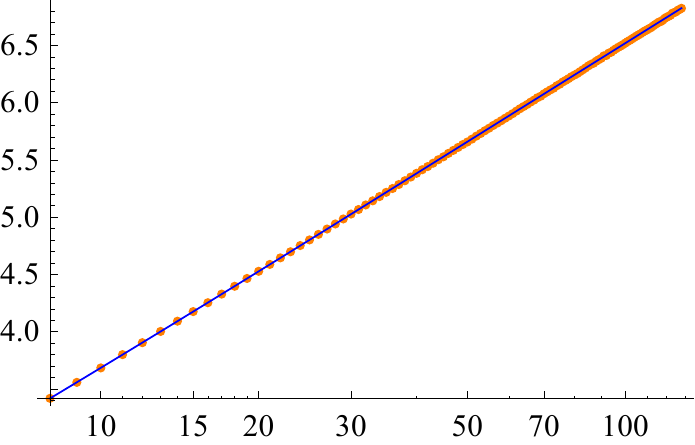}\rlap{\raisebox{2pt}{\hspace{-2pt}$L$}}
\hfill
\includegraphics[width=2.2in]{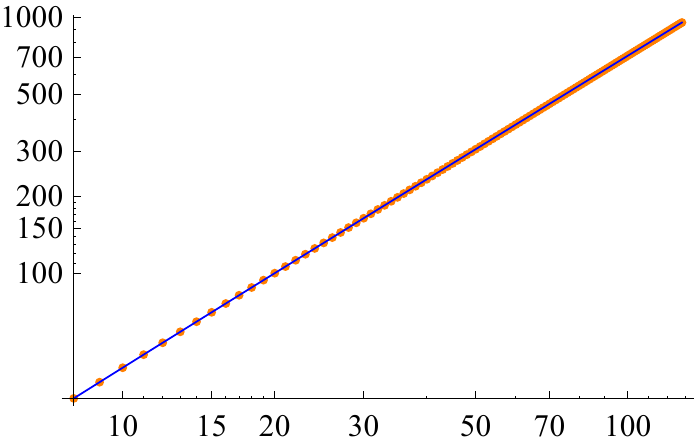}\rlap{\raisebox{2pt}{\hspace{-2pt}$L$}}
\hfill
\includegraphics[width=2.2in]{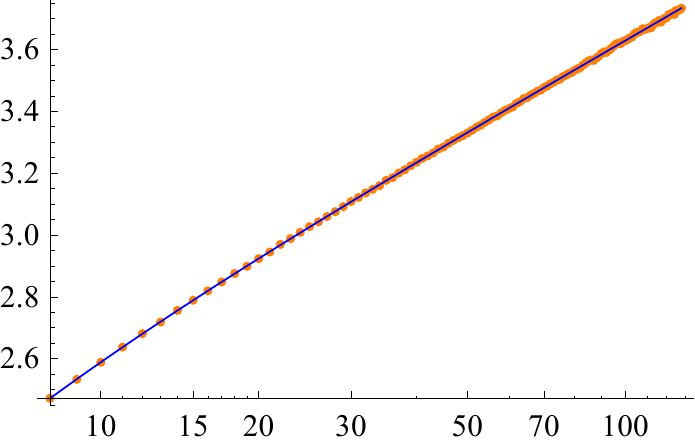}\rlap{\raisebox{2pt}{\hspace{-2pt}$L$}}
\end{center}
\vspace{-6pt}
\caption{
Active spanning tree simulation results with $y=1/2$ for various system sizes $L=8,9,\ldots,128$,
together with a curve fit.  The left
plot is a log-linear plot of the winding angle variance, and the fitted curve is $1.222\ln L +0.934-0.301(\ln L)^2/L+0.567(\ln L)/L-0.336/L$, which corresponds to the point $(0.5,1.222)$ in the winding angle variance coefficient plot in Fig.~\ref{Estimates}.
By comparison, the SLE winding coefficient prediction for $y=1/2$ is $1.245$.
The middle plot is a log-log plot of the expected estimator for the outer boundary length; the slope gives the dimension estimate in Fig.~\ref{Estimates}.  The right plot is a log-linear plot of the outer boundary's winding angle variance, whose slope gives the $y=1/2$ point in the right plot of Fig.~\ref{Estimates}.
}
\label{Estimates-y=0.5}
\end{figure*}

\begin{figure*}[t]
\vspace{6pt}
\begin{center}
\includegraphics[width=2.2in]{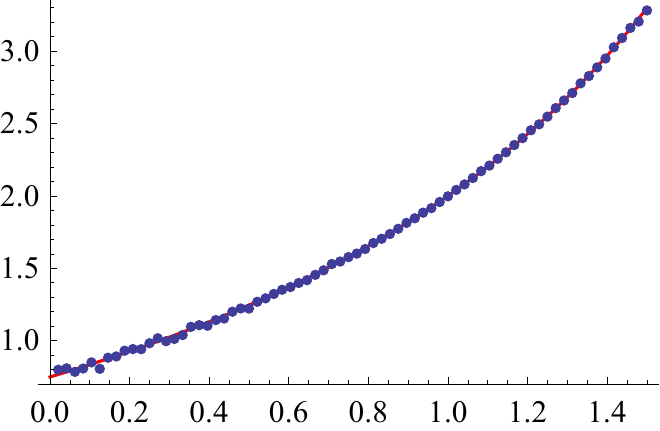}\rlap{\raisebox{5pt}{\hspace{-2pt}$y$}}
\hfill
\includegraphics[width=2.2in]{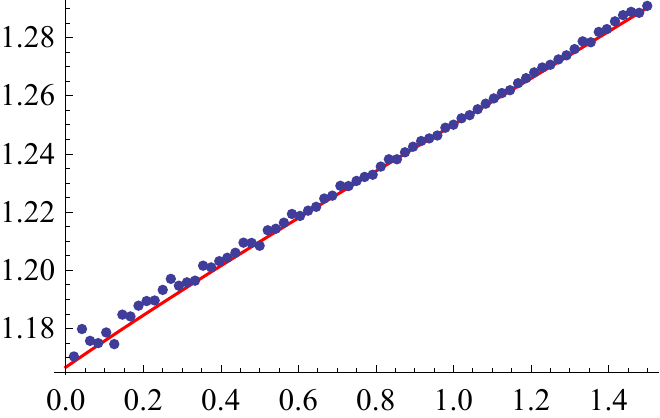}\rlap{\raisebox{5pt}{\hspace{-2pt}$y$}}
\hfill
\includegraphics[width=2.2in]{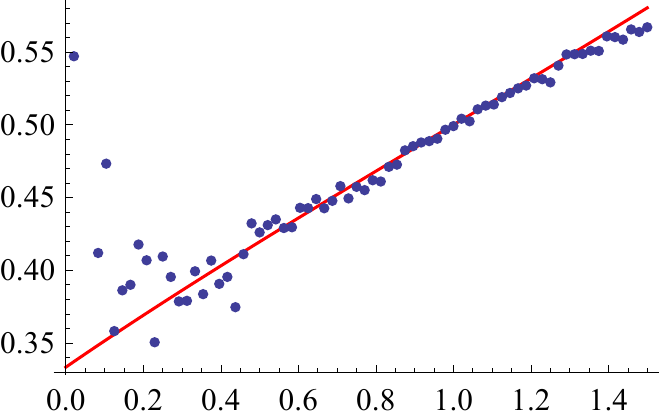}\rlap{\raisebox{5pt}{\hspace{-2pt}$y$}}
\end{center}
\vspace{-6pt}
\caption{Simulation estimates (points) and SLE predictions (curves) for 
the winding angle variance coefficient (left), outer boundary dimension (middle), and outer boundary winding angle variance coefficient (right)
of the active tree Peano curve, plotted as a function of the activity $y$.
The value of $\kappa'$ is determined by $y$ using Eqn.~\eqref{n-kappa} together with $n=y-1$.
The SLE predictions for these three quantities come from Eqns.~\eqref{wind-kappa}, \eqref{outer-dim-kappa}, and \eqref{outer-wind-kappa} respectively.}\label{Estimates}
\end{figure*}

\medskip
\bookmark[dest=sampling]{Sampling}
\hypertarget{sampling}{\textit{\textbf{Sampling.}}}
To test the conjecture, we sampled $y$-active spanning trees on an $L\times L$ region
of the square grid, for a variety of values of $y$ and $L$, and measured properties
of the Peano curve, comparing them to the values that would correspond to SLE,
as described in the next section.

We used a Markov chain to sample the $y$-active spanning trees.  The Markov chain picks
a random edge of the graph, and ``flips the edge'', adding it to the tree if it is not in the tree already, or removing the edge if it is in the tree.  This creates either a cycle or a dual cycle, which is broken by flipping another edge.  For each possible choice of the second edge to flip, we compute what the resulting activity would be, and choose the edge with probability proportional to $y$ to this activity.  The computations for each step of the Markov chain can be done in time which is typically proportional to the area of the cycle or dual cycle.  In the case $y=1$, this Markov chain has been proven to mix in polynomial time for any graph \cite{feder-mihail}.

Since there are no rigorous mixing time results except when $y=1$, we estimated the active-tree Markov chain's mixing rate by measuring decorrelation times of several properties of the active spanning tree.  Some of these measurements are illustrated in Figs.~\ref{fig:autocov-Lt} and~\ref{fig:autocov-yt}.  The autocovariance data appears to be consistent with polynomial time mixing for the 2D grid when $0<y\leq 3$.  We would not expect rapid mixing for the 2D grid when $y>3$ due to the first order phase transition of the associated critical FK model.

The active-tree Markov chain also provides a way to sample the FK model for any $q>0$ (recall the correspondence $y=1+\sqrt{q}$ and see Figs.~\ref{fig:autocov-Lt} and~\ref{fig:autocov-yt}).  For $q\ge 1$, the Swendsen--Wang algorithm~\cite{Swendsen-Wang}, the Wolff algorithm \cite{Wolff-PRL}, and the Chayes--Machta algorithm~\cite{Chayes-Machta} can be used for sampling, but for $q<1$, the only previous sampling algorithm is the single-bond heat-bath Markov chain~\cite{Sweeny,guo-jerrum:FK-bond,Sokal-Sportiello,Elci-Weigel}.  For $q>0$ ($y>1$) one could also alternate between the active spanning tree and FK representations, in a manner reminiscent of Swendsen--Wang's alternation between FK and spin representations.  Further tests should be carried out to compare these methods.

\medskip
\bookmark[dest=simulations]{Simulation results}
\hypertarget{simulations}{\textit{\textbf{Simulation results.}}}
We tested three properties of the active tree Peano curve to compare with $\SLE_{\kappa'}$.
These are (1) the variance of the winding angle of the Peano curve, which should scale as
\begin{equation} \label{wind-kappa}
4\kappa'/(\kappa'-4)^2\times\ln L
\end{equation} for space-filling $\SLE_{\kappa'}$ \cite{miller-sheffield:ig4},
(2) the dimension of the outer boundary of the Peano curve (i.e., the fractal dimension of the branches of the active tree), which should be 
\begin{equation} \label{outer-dim-kappa}
1+\kappa/8 = 1+2/\kappa'
\end{equation}
\cite{rohde-schramm,Beffara}, and (3) the winding angle variance of the outer boundary,
which should be
\begin{equation} \label{outer-wind-kappa}
4/\kappa' \times \ln L
\end{equation}
\cite{WW,duplantier}.
We did not measure the variation in the Loewner driving function because this is
difficult to do accurately, nor the
left-crossing probabilities \cite{Schramm:left},
because these would be affected by the values of $\rho_1$ and $\rho_2$ for the force points.

For each of many values of $y$, we produced samples for boxes with side length $L=8,9,\ldots,128$,
running the Markov chain between $4\times 10^5\times L^2$ and $10^6\times L^2$ steps,
collecting data on the active spanning tree Peano curve (Fig.~\ref{Estimates-y=0.5} shows the data for $y=0.5$).
We measured the variance of the winding of a random segment of the Peano curve, and the size
of the subtree or dual subtree rooted at a random edge, to measure the length of typical tree
branches.  We biased the random edge to be away from the boundary, according to the
square of the principal Dirichlet eigenvector, to better measure bulk
properties of the Peano curve. 
We used autocovariances in the sampled data (Figs.~\ref{fig:autocov-Lt}--\ref{fig:autocov-yt}) to estimate
the ``burn-in'' time for the Markov chain, and to estimate the uncertainty in the average
sampled data.

A study of the case $y=2$ (percolation) with much more data for many more $L$'s suggests
fitting to a function of the form $a \ln L + b + c\ln^2 L / L + d \ln L / L + e/L$, where the coefficient
of $\ln L$ is the desired quantity.
We then fit the measured variance in the winding of the curve and its outer boundary and the log of the mean subtree size to curves of this form.

The results of these curve fits, for each of many values of $y$, are shown in Fig.~\ref{Estimates}.
The estimates for the winding angle variance are nearly indistinguishable from the SLE prediction,
the estimates for the outer boundary dimension are quite close to the SLE prediction, and the
estimates for the outer boundary winding angle variance are noisy but not bad.  Overall
the experiments support the active-tree SLE conjecture.

\def\@rst #1 #2other{#1}
\newcommand\MR[1]{\relax\ifhmode\unskip\spacefactor3000 \space\fi
  \MRhref{\expandafter\@rst #1 other}{#1}}
\newcommand{\MRhref}[2]{\href{http://www.ams.org/mathscinet-getitem?mr=#1}{MR#1}}
\newcommand{\arXiv}[1]{\href{http://arxiv.org/abs/#1}{arXiv:#1}}
\newcommand{\arxiv}[1]{\href{http://arxiv.org/abs/#1}{#1}}

\bookmarksetup{startatroot}
\bibliography{activity}

\begin{thebibliography}{55}%
\makeatletter
\providecommand \@ifxundefined [1]{%
 \@ifx{#1\undefined}
}%
\providecommand \@ifnum [1]{%
 \ifnum #1\expandafter \@firstoftwo
 \else \expandafter \@secondoftwo
 \fi
}%
\providecommand \@ifx [1]{%
 \ifx #1\expandafter \@firstoftwo
 \else \expandafter \@secondoftwo
 \fi
}%
\providecommand \natexlab [1]{#1}%
\providecommand \enquote  [1]{``#1''}%
\providecommand \bibnamefont  [1]{#1}%
\providecommand \bibfnamefont [1]{#1}%
\providecommand \citenamefont [1]{#1}%
\providecommand \href@noop [0]{\@secondoftwo}%
\providecommand \href [0]{\begingroup \@sanitize@url \@href}%
\providecommand \@href[1]{\@@startlink{#1}\@@href}%
\providecommand \@@href[1]{\endgroup#1\@@endlink}%
\providecommand \@sanitize@url [0]{\catcode `\\12\catcode `\$12\catcode
  `\&12\catcode `\#12\catcode `\^12\catcode `\_12\catcode `\%12\relax}%
\providecommand \@@startlink[1]{}%
\providecommand \@@endlink[0]{}%
\providecommand \url  [0]{\begingroup\@sanitize@url \@url }%
\providecommand \@url [1]{\endgroup\@href {#1}{\urlprefix }}%
\providecommand \urlprefix  [0]{URL }%
\providecommand \Eprint [0]{\href }%
\providecommand \doibase [0]{http://dx.doi.org/}%
\providecommand \selectlanguage [0]{\@gobble}%
\providecommand \bibinfo  [0]{\@secondoftwo}%
\providecommand \bibfield  [0]{\@secondoftwo}%
\providecommand \translation [1]{[#1]}%
\providecommand \BibitemOpen [0]{}%
\providecommand \bibitemStop [0]{}%
\providecommand \bibitemNoStop [0]{.\EOS\space}%
\providecommand \EOS [0]{\spacefactor3000\relax}%
\providecommand \BibitemShut  [1]{\csname bibitem#1\endcsname}%
\let\auto@bib@innerbib\@empty
\bibitem [{\citenamefont {Fortuin}\ and\ \citenamefont {Kasteleyn}(1972)}]{FK}%
  \BibitemOpen
  \bibfield  {author} {\bibinfo {author} {\bibfnamefont {C.~M.}\ \bibnamefont
  {Fortuin}}\ and\ \bibinfo {author} {\bibfnamefont {P.~W.}\ \bibnamefont
  {Kasteleyn}},\ }\href {\doibase 10.1016/0031-8914(72)90045-6} {\bibfield
  {journal} {\bibinfo  {journal} {Physica}\ }\textbf {\bibinfo {volume} {57}},\
  \bibinfo {pages} {536} (\bibinfo {year} {1972})}\BibitemShut {NoStop}%
\bibitem [{\citenamefont {Baxter}(1971)}]{baxter}%
  \BibitemOpen
  \bibfield  {author} {\bibinfo {author} {\bibfnamefont {R.~J.}\ \bibnamefont
  {Baxter}},\ }\href@noop {} {\bibfield  {journal} {\bibinfo  {journal}
  {Studies in Appl.\ Math.}\ }\textbf {\bibinfo {volume} {50}},\ \bibinfo
  {pages} {51} (\bibinfo {year} {1971})}\BibitemShut {NoStop}%
\bibitem [{\citenamefont {Duminil-Copin}\ \emph {et~al.}(2015)\citenamefont
  {Duminil-Copin}, \citenamefont {Sidoravicius},\ and\ \citenamefont
  {Tassion}}]{DCST}%
  \BibitemOpen
  \bibfield  {author} {\bibinfo {author} {\bibfnamefont {H.}~\bibnamefont
  {Duminil-Copin}}, \bibinfo {author} {\bibfnamefont {V.}~\bibnamefont
  {Sidoravicius}}, \ and\ \bibinfo {author} {\bibfnamefont {V.}~\bibnamefont
  {Tassion}},\ }\href@noop {} {\  (\bibinfo {year} {2015})},\ \bibinfo {note}
  {\arXiv{1505.04159}}\BibitemShut {NoStop}%
\bibitem [{\citenamefont {Laanait}\ \emph {et~al.}(1986)\citenamefont
  {Laanait}, \citenamefont {Messager},\ and\ \citenamefont
  {Ruiz}}]{laanait-message-ruiz}%
  \BibitemOpen
  \bibfield  {author} {\bibinfo {author} {\bibfnamefont {L.}~\bibnamefont
  {Laanait}}, \bibinfo {author} {\bibfnamefont {A.}~\bibnamefont {Messager}}, \
  and\ \bibinfo {author} {\bibfnamefont {J.}~\bibnamefont {Ruiz}},\ }\href
  {http://projecteuclid.org/euclid.cmp/1104115499} {\bibfield  {journal}
  {\bibinfo  {journal} {Comm.\ Math.\ Phys.}\ }\textbf {\bibinfo {volume}
  {105}},\ \bibinfo {pages} {527} (\bibinfo {year} {1986})}\BibitemShut
  {NoStop}%
\bibitem [{\citenamefont {Beffara}\ and\ \citenamefont
  {Duminil-Copin}(2012)}]{Beffara-Duminil}%
  \BibitemOpen
  \bibfield  {author} {\bibinfo {author} {\bibfnamefont {V.}~\bibnamefont
  {Beffara}}\ and\ \bibinfo {author} {\bibfnamefont {H.}~\bibnamefont
  {Duminil-Copin}},\ }\href {\doibase 10.1007/s00440-011-0353-8} {\bibfield
  {journal} {\bibinfo  {journal} {Probab.\ Theory Related Fields}\ }\textbf
  {\bibinfo {volume} {153}},\ \bibinfo {pages} {511} (\bibinfo {year}
  {2012})}\BibitemShut {NoStop}%
\bibitem [{\citenamefont {Schramm}(2000)}]{Schramm:SLE}%
  \BibitemOpen
  \bibfield  {author} {\bibinfo {author} {\bibfnamefont {O.}~\bibnamefont
  {Schramm}},\ }\href {\doibase 10.1007/BF02803524} {\bibfield  {journal}
  {\bibinfo  {journal} {Israel J. Math.}\ }\textbf {\bibinfo {volume} {118}},\
  \bibinfo {pages} {221} (\bibinfo {year} {2000})}\BibitemShut {NoStop}%
\bibitem [{\citenamefont {Rohde}\ and\ \citenamefont
  {Schramm}(2005)}]{rohde-schramm}%
  \BibitemOpen
  \bibfield  {author} {\bibinfo {author} {\bibfnamefont {S.}~\bibnamefont
  {Rohde}}\ and\ \bibinfo {author} {\bibfnamefont {O.}~\bibnamefont
  {Schramm}},\ }\href {\doibase 10.4007/annals.2005.161.883} {\bibfield
  {journal} {\bibinfo  {journal} {Ann.\ of Math.~(2)}\ }\textbf {\bibinfo
  {volume} {161}},\ \bibinfo {pages} {883} (\bibinfo {year}
  {2005})}\BibitemShut {NoStop}%
\bibitem [{\citenamefont {Beffara}(2008)}]{Beffara}%
  \BibitemOpen
  \bibfield  {author} {\bibinfo {author} {\bibfnamefont {V.}~\bibnamefont
  {Beffara}},\ }\href {\doibase 10.1214/07-AOP364} {\bibfield  {journal}
  {\bibinfo  {journal} {Ann.\ Probab.}\ }\textbf {\bibinfo {volume} {36}},\
  \bibinfo {pages} {1421} (\bibinfo {year} {2008})}\BibitemShut {NoStop}%
\bibitem [{\citenamefont {Kager}\ and\ \citenamefont
  {Nienhuis}(2004)}]{kager-nienhuis:SLE}%
  \BibitemOpen
  \bibfield  {author} {\bibinfo {author} {\bibfnamefont {W.}~\bibnamefont
  {Kager}}\ and\ \bibinfo {author} {\bibfnamefont {B.}~\bibnamefont
  {Nienhuis}},\ }\href {\doibase 10.1023/B:JOSS.0000028058.87266.be} {\bibfield
   {journal} {\bibinfo  {journal} {J. Statist.\ Phys.}\ }\textbf {\bibinfo
  {volume} {115}},\ \bibinfo {pages} {1149} (\bibinfo {year}
  {2004})}\BibitemShut {NoStop}%
\bibitem [{\citenamefont {Schramm}(2007)}]{Schramm-ICM}%
  \BibitemOpen
  \bibfield  {author} {\bibinfo {author} {\bibfnamefont {O.}~\bibnamefont
  {Schramm}},\ }in\ \href {\doibase 10.4171/022-1/20} {\emph {\bibinfo
  {booktitle} {International {C}ongress of {M}athematicians. {V}ol.~{I}}}}\
  (\bibinfo  {publisher} {Eur.\ Math.\ Soc., Z\"urich},\ \bibinfo {year}
  {2007})\ pp.\ \bibinfo {pages} {513--543}\BibitemShut {NoStop}%
\bibitem [{\citenamefont {Miller}\ and\ \citenamefont
  {Sheffield}(2013)}]{miller-sheffield:ig4}%
  \BibitemOpen
  \bibfield  {author} {\bibinfo {author} {\bibfnamefont {J.}~\bibnamefont
  {Miller}}\ and\ \bibinfo {author} {\bibfnamefont {S.}~\bibnamefont
  {Sheffield}},\ }\href@noop {} {\  (\bibinfo {year} {2013})},\ \bibinfo {note}
  {\arXiv{1302.4738}}\BibitemShut {NoStop}%
\bibitem [{\citenamefont {Chelkak}\ \emph {et~al.}(2014)\citenamefont
  {Chelkak}, \citenamefont {Duminil-Copin}, \citenamefont {Hongler},
  \citenamefont {Kemppainen},\ and\ \citenamefont {Smirnov}}]{CDCHKS:ising}%
  \BibitemOpen
  \bibfield  {author} {\bibinfo {author} {\bibfnamefont {D.}~\bibnamefont
  {Chelkak}}, \bibinfo {author} {\bibfnamefont {H.}~\bibnamefont
  {Duminil-Copin}}, \bibinfo {author} {\bibfnamefont {C.}~\bibnamefont
  {Hongler}}, \bibinfo {author} {\bibfnamefont {A.}~\bibnamefont {Kemppainen}},
  \ and\ \bibinfo {author} {\bibfnamefont {S.}~\bibnamefont {Smirnov}},\ }\href
  {\doibase 10.1016/j.crma.2013.12.002} {\bibfield  {journal} {\bibinfo
  {journal} {C. R. Math.\ Acad.\ Sci.\ Paris}\ }\textbf {\bibinfo {volume}
  {352}},\ \bibinfo {pages} {157} (\bibinfo {year} {2014})}\BibitemShut
  {NoStop}%
\bibitem [{\citenamefont {Kemppainen}\ and\ \citenamefont
  {Smirnov}(2015)}]{kemppainen-smirnov}%
  \BibitemOpen
  \bibfield  {author} {\bibinfo {author} {\bibfnamefont {A.}~\bibnamefont
  {Kemppainen}}\ and\ \bibinfo {author} {\bibfnamefont {S.}~\bibnamefont
  {Smirnov}},\ }\href@noop {} {\  (\bibinfo {year} {2015})},\ \bibinfo {note}
  {\arXiv{1509.08858}}\BibitemShut {NoStop}%
\bibitem [{\citenamefont {Lawler}\ \emph {et~al.}(2004)\citenamefont {Lawler},
  \citenamefont {Schramm},\ and\ \citenamefont {Werner}}]{LSW:UST}%
  \BibitemOpen
  \bibfield  {author} {\bibinfo {author} {\bibfnamefont {G.~F.}\ \bibnamefont
  {Lawler}}, \bibinfo {author} {\bibfnamefont {O.}~\bibnamefont {Schramm}}, \
  and\ \bibinfo {author} {\bibfnamefont {W.}~\bibnamefont {Werner}},\ }\href
  {\doibase 10.1214/aop/1079021469} {\bibfield  {journal} {\bibinfo  {journal}
  {Ann.\ Probab.}\ }\textbf {\bibinfo {volume} {32}},\ \bibinfo {pages} {939}
  (\bibinfo {year} {2004})}\BibitemShut {NoStop}%
\bibitem [{\citenamefont {Bl{\"o}te}\ and\ \citenamefont
  {Nienhuis}(1994)}]{blote-nienhuis}%
  \BibitemOpen
  \bibfield  {author} {\bibinfo {author} {\bibfnamefont {H.~W.~J.}\
  \bibnamefont {Bl{\"o}te}}\ and\ \bibinfo {author} {\bibfnamefont
  {B.}~\bibnamefont {Nienhuis}},\ }\href {\doibase 10.1103/PhysRevLett.72.1372}
  {\bibfield  {journal} {\bibinfo  {journal} {Phys.\ Rev.\ Lett.}\ }\textbf
  {\bibinfo {volume} {72}},\ \bibinfo {pages} {1372} (\bibinfo {year}
  {1994})}\BibitemShut {NoStop}%
\bibitem [{\citenamefont {Kondev}\ \emph {et~al.}(1996)\citenamefont {Kondev},
  \citenamefont {de~Gier},\ and\ \citenamefont {Nienhuis}}]{KdGN}%
  \BibitemOpen
  \bibfield  {author} {\bibinfo {author} {\bibfnamefont {J.}~\bibnamefont
  {Kondev}}, \bibinfo {author} {\bibfnamefont {J.}~\bibnamefont {de~Gier}}, \
  and\ \bibinfo {author} {\bibfnamefont {B.}~\bibnamefont {Nienhuis}},\ }\href
  {\doibase 10.1088/0305-4470/29/20/007} {\bibfield  {journal} {\bibinfo
  {journal} {J. Phys.\ A}\ }\textbf {\bibinfo {volume} {29}},\ \bibinfo {pages}
  {6489} (\bibinfo {year} {1996})}\BibitemShut {NoStop}%
\bibitem [{\citenamefont {Batchelor}\ \emph {et~al.}(1996)\citenamefont
  {Batchelor}, \citenamefont {Bl\"ote}, \citenamefont {Nienhuis},\ and\
  \citenamefont {Yung}}]{BBNY}%
  \BibitemOpen
  \bibfield  {author} {\bibinfo {author} {\bibfnamefont {M.~T.}\ \bibnamefont
  {Batchelor}}, \bibinfo {author} {\bibfnamefont {H.~W.~J.}\ \bibnamefont
  {Bl\"ote}}, \bibinfo {author} {\bibfnamefont {B.}~\bibnamefont {Nienhuis}}, \
  and\ \bibinfo {author} {\bibfnamefont {C.~M.}\ \bibnamefont {Yung}},\ }\href
  {\doibase 10.1088/0305-4470/29/16/001} {\bibfield  {journal} {\bibinfo
  {journal} {J. Phys.\ A}\ }\textbf {\bibinfo {volume} {29}},\ \bibinfo {pages}
  {L399} (\bibinfo {year} {1996})}\BibitemShut {NoStop}%
\bibitem [{\citenamefont {Jacobsen}(1999)}]{jacobsen:compact}%
  \BibitemOpen
  \bibfield  {author} {\bibinfo {author} {\bibfnamefont {J.~L.}\ \bibnamefont
  {Jacobsen}},\ }\href {\doibase 10.1088/0305-4470/32/29/305} {\bibfield
  {journal} {\bibinfo  {journal} {J. Phys.~A}\ }\textbf {\bibinfo {volume}
  {32}},\ \bibinfo {pages} {5445} (\bibinfo {year} {1999})}\BibitemShut
  {NoStop}%
\bibitem [{\citenamefont {Daryaei}\ \emph {et~al.}(2012)\citenamefont
  {Daryaei}, \citenamefont {Ara\'ujo}, \citenamefont {Schrenk}, \citenamefont
  {Rouhani},\ and\ \citenamefont {Herrmann}}]{watersheds-sle}%
  \BibitemOpen
  \bibfield  {author} {\bibinfo {author} {\bibfnamefont {E.}~\bibnamefont
  {Daryaei}}, \bibinfo {author} {\bibfnamefont {N.~A.~M.}\ \bibnamefont
  {Ara\'ujo}}, \bibinfo {author} {\bibfnamefont {K.~J.}\ \bibnamefont
  {Schrenk}}, \bibinfo {author} {\bibfnamefont {S.}~\bibnamefont {Rouhani}}, \
  and\ \bibinfo {author} {\bibfnamefont {H.~J.}\ \bibnamefont {Herrmann}},\
  }\href {\doibase 10.1103/PhysRevLett.109.218701} {\bibfield  {journal}
  {\bibinfo  {journal} {Phys.\ Rev.\ Lett.}\ }\textbf {\bibinfo {volume}
  {109}},\ \bibinfo {pages} {218701} (\bibinfo {year} {2012})}\BibitemShut
  {NoStop}%
\bibitem [{\citenamefont {Wieland}\ and\ \citenamefont {Wilson}(2003)}]{WW}%
  \BibitemOpen
  \bibfield  {author} {\bibinfo {author} {\bibfnamefont {B.}~\bibnamefont
  {Wieland}}\ and\ \bibinfo {author} {\bibfnamefont {D.~B.}\ \bibnamefont
  {Wilson}},\ }\href {\doibase 10.1103/PhysRevE.68.056101} {\bibfield
  {journal} {\bibinfo  {journal} {Phys.\ Rev.\ E}\ }\textbf {\bibinfo {volume}
  {68}},\ \bibinfo {pages} {056101} (\bibinfo {year} {2003})}\BibitemShut
  {NoStop}%
\bibitem [{\citenamefont {Wilson}(2004)}]{wilson:RGB}%
  \BibitemOpen
  \bibfield  {author} {\bibinfo {author} {\bibfnamefont {D.~B.}\ \bibnamefont
  {Wilson}},\ }\href {\doibase 10.1103/PhysRevE.69.037105} {\bibfield
  {journal} {\bibinfo  {journal} {Phys.\ Rev.\ E}\ }\textbf {\bibinfo {volume}
  {69}},\ \bibinfo {pages} {037105} (\bibinfo {year} {2004})}\BibitemShut
  {NoStop}%
\bibitem [{\citenamefont {Nienhuis}(1982)}]{Nienhuis-82}%
  \BibitemOpen
  \bibfield  {author} {\bibinfo {author} {\bibfnamefont {B.}~\bibnamefont
  {Nienhuis}},\ }\href {\doibase 10.1103/PhysRevLett.49.1062} {\bibfield
  {journal} {\bibinfo  {journal} {Phys.\ Rev.\ Lett.}\ }\textbf {\bibinfo
  {volume} {49}},\ \bibinfo {pages} {1062} (\bibinfo {year}
  {1982})}\BibitemShut {NoStop}%
\bibitem [{\citenamefont {Cardy}(2005)}]{Cardy}%
  \BibitemOpen
  \bibfield  {author} {\bibinfo {author} {\bibfnamefont {J.}~\bibnamefont
  {Cardy}},\ }\href {\doibase 10.1016/j.aop.2005.04.001} {\bibfield  {journal}
  {\bibinfo  {journal} {Ann.\ Physics}\ }\textbf {\bibinfo {volume} {318}},\
  \bibinfo {pages} {81} (\bibinfo {year} {2005})}\BibitemShut {NoStop}%
\bibitem [{\citenamefont {Tutte}(1954)}]{Tutte}%
  \BibitemOpen
  \bibfield  {author} {\bibinfo {author} {\bibfnamefont {W.~T.}\ \bibnamefont
  {Tutte}},\ }\href@noop {} {\bibfield  {journal} {\bibinfo  {journal} {Canad.\
  J. Math.}\ }\textbf {\bibinfo {volume} {6}},\ \bibinfo {pages} {80} (\bibinfo
  {year} {1954})}\BibitemShut {NoStop}%
\bibitem [{\citenamefont {Bernardi}(2008{\natexlab{a}})}]{Bernardi0}%
  \BibitemOpen
  \bibfield  {author} {\bibinfo {author} {\bibfnamefont {O.}~\bibnamefont
  {Bernardi}},\ }\href {\doibase 10.1007/s00026-008-0343-4} {\bibfield
  {journal} {\bibinfo  {journal} {Ann.\ Comb.}\ }\textbf {\bibinfo {volume}
  {12}},\ \bibinfo {pages} {139} (\bibinfo {year}
  {2008}{\natexlab{a}})}\BibitemShut {NoStop}%
\bibitem [{\citenamefont {Bernardi}(2008{\natexlab{b}})}]{Bernardi}%
  \BibitemOpen
  \bibfield  {author} {\bibinfo {author} {\bibfnamefont {O.}~\bibnamefont
  {Bernardi}},\ }\href
  {http://www.combinatorics.org/Volume_15/Abstracts/v15i1r109.html} {\bibfield
  {journal} {\bibinfo  {journal} {Electron.\ J. Combin.}\ }\textbf {\bibinfo
  {volume} {15}},\ \bibinfo {pages} {Paper 109} (\bibinfo {year}
  {2008}{\natexlab{b}})}\BibitemShut {NoStop}%
\bibitem [{\citenamefont {Courtiel}(2014)}]{Courtiel}%
  \BibitemOpen
  \bibfield  {author} {\bibinfo {author} {\bibfnamefont {J.}~\bibnamefont
  {Courtiel}},\ }\href@noop {} {\  (\bibinfo {year} {2014})},\ \bibinfo {note}
  {\arXiv{1412.2081}}\BibitemShut {NoStop}%
\bibitem [{\citenamefont {Sheffield}(2009)}]{Sheffield-exploration}%
  \BibitemOpen
  \bibfield  {author} {\bibinfo {author} {\bibfnamefont {S.}~\bibnamefont
  {Sheffield}},\ }\href {\doibase 10.1215/00127094-2009-007} {\bibfield
  {journal} {\bibinfo  {journal} {Duke Math.\ J.}\ }\textbf {\bibinfo {volume}
  {147}},\ \bibinfo {pages} {79} (\bibinfo {year} {2009})}\BibitemShut
  {NoStop}%
\bibitem [{\citenamefont
  {Sheffield}(pear{\natexlab{a}})}]{Sheffield-HCbijection}%
  \BibitemOpen
  \bibfield  {author} {\bibinfo {author} {\bibfnamefont {S.}~\bibnamefont
  {Sheffield}},\ }\href@noop {} {\bibfield  {journal} {\bibinfo  {journal}
  {Ann.\ Probab.}\ } (\bibinfo {year} {to appear}{\natexlab{a}})},\ \bibinfo
  {note} {\arXiv{1108.2241}}\BibitemShut {NoStop}%
\bibitem [{\citenamefont {Baxter}\ \emph {et~al.}(1976)\citenamefont {Baxter},
  \citenamefont {Kelland},\ and\ \citenamefont {Wu}}]{BKW}%
  \BibitemOpen
  \bibfield  {author} {\bibinfo {author} {\bibfnamefont {R.~J.}\ \bibnamefont
  {Baxter}}, \bibinfo {author} {\bibfnamefont {S.~B.}\ \bibnamefont {Kelland}},
  \ and\ \bibinfo {author} {\bibfnamefont {F.~Y.}\ \bibnamefont {Wu}},\ }\href
  {\doibase 10.1088/0305-4470/9/3/009} {\bibfield  {journal} {\bibinfo
  {journal} {J. Phys.\ A}\ }\textbf {\bibinfo {volume} {9}},\ \bibinfo {pages}
  {397} (\bibinfo {year} {1976})}\BibitemShut {NoStop}%
\bibitem [{\citenamefont {Jaeger}(1988)}]{Jaeger-88}%
  \BibitemOpen
  \bibfield  {author} {\bibinfo {author} {\bibfnamefont {F.}~\bibnamefont
  {Jaeger}},\ }\href {\doibase 10.1016/0095-8956(88)90083-4} {\bibfield
  {journal} {\bibinfo  {journal} {J. Combin.\ Theory Ser.~B}\ }\textbf
  {\bibinfo {volume} {44}},\ \bibinfo {pages} {127} (\bibinfo {year}
  {1988})}\BibitemShut {NoStop}%
\bibitem [{\citenamefont {Dub{\'e}dat}(2009{\natexlab{a}})}]{dubedat:SLE-GFF}%
  \BibitemOpen
  \bibfield  {author} {\bibinfo {author} {\bibfnamefont {J.}~\bibnamefont
  {Dub{\'e}dat}},\ }\href {\doibase 10.1090/S0894-0347-09-00636-5} {\bibfield
  {journal} {\bibinfo  {journal} {J. Amer.\ Math.\ Soc.}\ }\textbf {\bibinfo
  {volume} {22}},\ \bibinfo {pages} {995} (\bibinfo {year}
  {2009}{\natexlab{a}})}\BibitemShut {NoStop}%
\bibitem [{\citenamefont {Sheffield}(pear{\natexlab{b}})}]{sheffield:welding}%
  \BibitemOpen
  \bibfield  {author} {\bibinfo {author} {\bibfnamefont {S.}~\bibnamefont
  {Sheffield}},\ }\href@noop {} {\bibfield  {journal} {\bibinfo  {journal}
  {Ann.\ Probab.}\ } (\bibinfo {year} {to appear}{\natexlab{b}})},\ \bibinfo
  {note} {\arXiv{1012.4797}}\BibitemShut {NoStop}%
\bibitem [{\citenamefont {Miller}\ and\ \citenamefont
  {Sheffield}(pear)}]{miller-sheffield:ig1}%
  \BibitemOpen
  \bibfield  {author} {\bibinfo {author} {\bibfnamefont {J.}~\bibnamefont
  {Miller}}\ and\ \bibinfo {author} {\bibfnamefont {S.}~\bibnamefont
  {Sheffield}},\ }\href@noop {} {\bibfield  {journal} {\bibinfo  {journal}
  {Probab.\ Theory Related Fields}\ } (\bibinfo {year} {to appear})},\ \bibinfo
  {note} {\arXiv{1201.1496}}\BibitemShut {NoStop}%
\bibitem [{\citenamefont {Zhan}(2008)}]{Zhan}%
  \BibitemOpen
  \bibfield  {author} {\bibinfo {author} {\bibfnamefont {D.}~\bibnamefont
  {Zhan}},\ }\href {\doibase 10.1007/s00222-008-0132-z} {\bibfield  {journal}
  {\bibinfo  {journal} {Invent.\ Math.}\ }\textbf {\bibinfo {volume} {174}},\
  \bibinfo {pages} {309} (\bibinfo {year} {2008})}\BibitemShut {NoStop}%
\bibitem [{\citenamefont {Dub{\'e}dat}(2009{\natexlab{b}})}]{Dubedat}%
  \BibitemOpen
  \bibfield  {author} {\bibinfo {author} {\bibfnamefont {J.}~\bibnamefont
  {Dub{\'e}dat}},\ }\href@noop {} {\bibfield  {journal} {\bibinfo  {journal}
  {Ann.\ Sci.\ \'Ec.\ Norm.\ Sup\'er.\ (4)}\ }\textbf {\bibinfo {volume}
  {42}},\ \bibinfo {pages} {697} (\bibinfo {year}
  {2009}{\natexlab{b}})}\BibitemShut {NoStop}%
\bibitem [{\citenamefont {Gwynne}\ \emph {et~al.}(2016)\citenamefont {Gwynne},
  \citenamefont {Kassel}, \citenamefont {Miller},\ and\ \citenamefont
  {Wilson}}]{GKMW:active-tree-map}%
  \BibitemOpen
  \bibfield  {author} {\bibinfo {author} {\bibfnamefont {E.}~\bibnamefont
  {Gwynne}}, \bibinfo {author} {\bibfnamefont {A.}~\bibnamefont {Kassel}},
  \bibinfo {author} {\bibfnamefont {J.}~\bibnamefont {Miller}}, \ and\ \bibinfo
  {author} {\bibfnamefont {D.~B.}\ \bibnamefont {Wilson}},\ }\href@noop {} {\
  (\bibinfo {year} {2016})},\ \bibinfo {note} {\arXiv{1603.09722}}\BibitemShut
  {NoStop}%
\bibitem [{\citenamefont {{Duplantier}}\ \emph {et~al.}(2014)\citenamefont
  {{Duplantier}}, \citenamefont {{Miller}},\ and\ \citenamefont
  {{Sheffield}}}]{DMS:mating}%
  \BibitemOpen
  \bibfield  {author} {\bibinfo {author} {\bibfnamefont {B.}~\bibnamefont
  {{Duplantier}}}, \bibinfo {author} {\bibfnamefont {J.}~\bibnamefont
  {{Miller}}}, \ and\ \bibinfo {author} {\bibfnamefont {S.}~\bibnamefont
  {{Sheffield}}},\ }\href@noop {} {\  (\bibinfo {year} {2014})},\ \bibinfo
  {note} {\arXiv{1409.7055}}\BibitemShut {NoStop}%
\bibitem [{\citenamefont {Miller}\ and\ \citenamefont
  {Sheffield}(2015)}]{miller-sheffield-peano}%
  \BibitemOpen
  \bibfield  {author} {\bibinfo {author} {\bibfnamefont {J.}~\bibnamefont
  {Miller}}\ and\ \bibinfo {author} {\bibfnamefont {S.}~\bibnamefont
  {Sheffield}},\ }\href@noop {} {\  (\bibinfo {year} {2015})},\ \bibinfo {note}
  {\arXiv{1506.03804}}\BibitemShut {NoStop}%
\bibitem [{\citenamefont {Gwynne}\ \emph
  {et~al.}(2015{\natexlab{a}})\citenamefont {Gwynne}, \citenamefont {Holden},
  \citenamefont {Miller},\ and\ \citenamefont {Sun}}]{GHMS:covariance}%
  \BibitemOpen
  \bibfield  {author} {\bibinfo {author} {\bibfnamefont {E.}~\bibnamefont
  {Gwynne}}, \bibinfo {author} {\bibfnamefont {N.}~\bibnamefont {Holden}},
  \bibinfo {author} {\bibfnamefont {J.}~\bibnamefont {Miller}}, \ and\ \bibinfo
  {author} {\bibfnamefont {X.}~\bibnamefont {Sun}},\ }\href@noop {} {\
  (\bibinfo {year} {2015}{\natexlab{a}})},\ \bibinfo {note}
  {\arXiv{1510.04687}}\BibitemShut {NoStop}%
\bibitem [{\citenamefont {Gwynne}\ \emph
  {et~al.}(2015{\natexlab{b}})\citenamefont {Gwynne}, \citenamefont {Mao},\
  and\ \citenamefont {Sun}}]{gwynne-mao-sun}%
  \BibitemOpen
  \bibfield  {author} {\bibinfo {author} {\bibfnamefont {E.}~\bibnamefont
  {Gwynne}}, \bibinfo {author} {\bibfnamefont {C.}~\bibnamefont {Mao}}, \ and\
  \bibinfo {author} {\bibfnamefont {X.}~\bibnamefont {Sun}},\ }\href@noop {} {\
   (\bibinfo {year} {2015}{\natexlab{b}})},\ \bibinfo {note}
  {\arXiv{1502.00546}}\BibitemShut {NoStop}%
\bibitem [{\citenamefont {Gwynne}\ and\ \citenamefont
  {Sun}(2015{\natexlab{a}})}]{gwynne-sun-II}%
  \BibitemOpen
  \bibfield  {author} {\bibinfo {author} {\bibfnamefont {E.}~\bibnamefont
  {Gwynne}}\ and\ \bibinfo {author} {\bibfnamefont {X.}~\bibnamefont {Sun}},\
  }\href@noop {} {\  (\bibinfo {year} {2015}{\natexlab{a}})},\ \bibinfo {note}
  {\arXiv{1505.03375}}\BibitemShut {NoStop}%
\bibitem [{\citenamefont {Gwynne}\ and\ \citenamefont
  {Sun}(2015{\natexlab{b}})}]{gwynne-sun-III}%
  \BibitemOpen
  \bibfield  {author} {\bibinfo {author} {\bibfnamefont {E.}~\bibnamefont
  {Gwynne}}\ and\ \bibinfo {author} {\bibfnamefont {X.}~\bibnamefont {Sun}},\
  }\href@noop {} {\  (\bibinfo {year} {2015}{\natexlab{b}})},\ \bibinfo {note}
  {\arXiv{1510.06346}}\BibitemShut {NoStop}%
\bibitem [{\citenamefont {Gioan}\ and\ \citenamefont
  {Las~Vergnas}(2005)}]{gioan-vergnas}%
  \BibitemOpen
  \bibfield  {author} {\bibinfo {author} {\bibfnamefont {E.}~\bibnamefont
  {Gioan}}\ and\ \bibinfo {author} {\bibfnamefont {M.}~\bibnamefont
  {Las~Vergnas}},\ }\href {\doibase 10.1016/j.disc.2005.04.010} {\bibfield
  {journal} {\bibinfo  {journal} {Discrete Math.}\ }\textbf {\bibinfo {volume}
  {298}},\ \bibinfo {pages} {169} (\bibinfo {year} {2005})}\BibitemShut
  {NoStop}%
\bibitem [{\citenamefont {Kenyon}\ \emph {et~al.}(2015)\citenamefont {Kenyon},
  \citenamefont {Miller}, \citenamefont {Sheffield},\ and\ \citenamefont
  {Wilson}}]{KMSW1}%
  \BibitemOpen
  \bibfield  {author} {\bibinfo {author} {\bibfnamefont {R.~W.}\ \bibnamefont
  {Kenyon}}, \bibinfo {author} {\bibfnamefont {J.}~\bibnamefont {Miller}},
  \bibinfo {author} {\bibfnamefont {S.}~\bibnamefont {Sheffield}}, \ and\
  \bibinfo {author} {\bibfnamefont {D.~B.}\ \bibnamefont {Wilson}},\
  }\href@noop {} {\  (\bibinfo {year} {2015})},\ \bibinfo {note}
  {\arXiv{1511.04068}}\BibitemShut {NoStop}%
\bibitem [{\citenamefont {Feder}\ and\ \citenamefont
  {Mihail}(1992)}]{feder-mihail}%
  \BibitemOpen
  \bibfield  {author} {\bibinfo {author} {\bibfnamefont {T.}~\bibnamefont
  {Feder}}\ and\ \bibinfo {author} {\bibfnamefont {M.}~\bibnamefont {Mihail}},\
  }in\ \href {\doibase 10.1145/129712.129716} {\emph {\bibinfo {booktitle}
  {Proc.\ 24th ACM Symp.\ Theory of Computing}}}\ (\bibinfo {year} {1992})\
  pp.\ \bibinfo {pages} {26--38}\BibitemShut {NoStop}%
\bibitem [{\citenamefont {Swendsen}\ and\ \citenamefont
  {Wang}(1987)}]{Swendsen-Wang}%
  \BibitemOpen
  \bibfield  {author} {\bibinfo {author} {\bibfnamefont {R.~H.}\ \bibnamefont
  {Swendsen}}\ and\ \bibinfo {author} {\bibfnamefont {J.-S.}\ \bibnamefont
  {Wang}},\ }\href {\doibase 10.1103/PhysRevLett.58.86} {\bibfield  {journal}
  {\bibinfo  {journal} {Phys.\ Rev.\ Lett.}\ }\textbf {\bibinfo {volume}
  {58}},\ \bibinfo {pages} {86} (\bibinfo {year} {1987})}\BibitemShut {NoStop}%
\bibitem [{\citenamefont {Wolff}(1989)}]{Wolff-PRL}%
  \BibitemOpen
  \bibfield  {author} {\bibinfo {author} {\bibfnamefont {U.}~\bibnamefont
  {Wolff}},\ }\href {\doibase 10.1103/PhysRevLett.62.361} {\bibfield  {journal}
  {\bibinfo  {journal} {Phys.\ Rev.\ Lett.}\ }\textbf {\bibinfo {volume}
  {62}},\ \bibinfo {pages} {361} (\bibinfo {year} {1989})}\BibitemShut
  {NoStop}%
\bibitem [{\citenamefont {Chayes}\ and\ \citenamefont
  {Machta}(1998)}]{Chayes-Machta}%
  \BibitemOpen
  \bibfield  {author} {\bibinfo {author} {\bibfnamefont {L.}~\bibnamefont
  {Chayes}}\ and\ \bibinfo {author} {\bibfnamefont {J.}~\bibnamefont
  {Machta}},\ }\href {\doibase 10.1016/S0378-4371(97)00637-7} {\bibfield
  {journal} {\bibinfo  {journal} {Physica A}\ }\textbf {\bibinfo {volume}
  {254}},\ \bibinfo {pages} {477} (\bibinfo {year} {1998})}\BibitemShut
  {NoStop}%
\bibitem [{\citenamefont {Sweeny}(1983)}]{Sweeny}%
  \BibitemOpen
  \bibfield  {author} {\bibinfo {author} {\bibfnamefont {M.}~\bibnamefont
  {Sweeny}},\ }\href {\doibase 10.1103/PhysRevB.27.4445} {\bibfield  {journal}
  {\bibinfo  {journal} {Phys.\ Rev.~B}\ }\textbf {\bibinfo {volume} {27}},\
  \bibinfo {pages} {4445} (\bibinfo {year} {1983})}\BibitemShut {NoStop}%
\bibitem [{\citenamefont {Guo}\ and\ \citenamefont
  {Jerrum}(2016)}]{guo-jerrum:FK-bond}%
  \BibitemOpen
  \bibfield  {author} {\bibinfo {author} {\bibfnamefont {H.}~\bibnamefont
  {Guo}}\ and\ \bibinfo {author} {\bibfnamefont {M.}~\bibnamefont {Jerrum}},\
  }\href@noop {} {\  (\bibinfo {year} {2016})},\ \bibinfo {note}
  {\arXiv{1605.00139}}\BibitemShut {NoStop}%
\bibitem [{\citenamefont {Deng}\ \emph {et~al.}(2010)\citenamefont {Deng},
  \citenamefont {Zhang}, \citenamefont {Garoni}, \citenamefont {Sokal},\ and\
  \citenamefont {Sportiello}}]{Sokal-Sportiello}%
  \BibitemOpen
  \bibfield  {author} {\bibinfo {author} {\bibfnamefont {Y.}~\bibnamefont
  {Deng}}, \bibinfo {author} {\bibfnamefont {W.}~\bibnamefont {Zhang}},
  \bibinfo {author} {\bibfnamefont {T.~M.}\ \bibnamefont {Garoni}}, \bibinfo
  {author} {\bibfnamefont {A.~D.}\ \bibnamefont {Sokal}}, \ and\ \bibinfo
  {author} {\bibfnamefont {A.}~\bibnamefont {Sportiello}},\ }\href {\doibase
  10.1103/PhysRevE.81.020102} {\bibfield  {journal} {\bibinfo  {journal}
  {Phys.\ Rev.\ E}\ }\textbf {\bibinfo {volume} {81}},\ \bibinfo {pages}
  {020102} (\bibinfo {year} {2010})}\BibitemShut {NoStop}%
\bibitem [{\citenamefont {El\c{c}i}\ and\ \citenamefont
  {Weigel}(2013)}]{Elci-Weigel}%
  \BibitemOpen
  \bibfield  {author} {\bibinfo {author} {\bibfnamefont {E.~M.}\ \bibnamefont
  {El\c{c}i}}\ and\ \bibinfo {author} {\bibfnamefont {M.}~\bibnamefont
  {Weigel}},\ }\href {\doibase 10.1103/PhysRevE.88.033303} {\bibfield
  {journal} {\bibinfo  {journal} {Phys.\ Rev.\ E}\ }\textbf {\bibinfo {volume}
  {88}},\ \bibinfo {pages} {033303} (\bibinfo {year} {2013})}\BibitemShut
  {NoStop}%
\bibitem [{\citenamefont {Duplantier}(2004)}]{duplantier}%
  \BibitemOpen
  \bibfield  {author} {\bibinfo {author} {\bibfnamefont {B.}~\bibnamefont
  {Duplantier}},\ }\href@noop {} {\bibfield  {journal} {\bibinfo  {journal}
  {Proc.\ Sympos.\ Pure Math.}\ }\textbf {\bibinfo {volume} {72}},\ \bibinfo
  {pages} {365} (\bibinfo {year} {2004})}\BibitemShut {NoStop}%
\bibitem [{\citenamefont {Schramm}(2001)}]{Schramm:left}%
  \BibitemOpen
  \bibfield  {author} {\bibinfo {author} {\bibfnamefont {O.}~\bibnamefont
  {Schramm}},\ }\href {\doibase 10.1214/ECP.v6-1041} {\bibfield  {journal}
  {\bibinfo  {journal} {Electron.\ Comm.\ Probab.}\ }\textbf {\bibinfo {volume}
  {6}},\ \bibinfo {pages} {115} (\bibinfo {year} {2001})}\BibitemShut {NoStop}%
\end{thebibliography}%

\end{document}